\documentclass[11pt,a4paper]{article}
\usepackage[T1]{fontenc}
\usepackage{lmodern,amsmath,amssymb,amsthm}
\usepackage[a4paper,left=26mm,right=26mm,top=24mm,bottom=24mm]{geometry}
\usepackage{xcolor,hyperref}
\definecolor{linkink}{RGB}{24,55,75}
\hypersetup{colorlinks=true,linkcolor=linkink,citecolor=linkink,urlcolor=linkink,
 pdftitle={Prime Partitions of Matroids and Matroid Covering Games},
 pdfauthor={Pengfei Liu, Han Xiao, Qizhi Fang},pdfsubject={Parametric matroid partitions and the nucleolus}}
\theoremstyle{plain}
\newtheorem{theorem}{Theorem}[section]
\newtheorem{lemma}[theorem]{Lemma}

\newtheorem{corollary}[theorem]{Corollary}
\theoremstyle{definition}
\newtheorem{definition}[theorem]{Definition}
\newtheorem{example}[theorem]{Example}
\theoremstyle{remark}

\numberwithin{equation}{section}
\title{\bfseries Prime Partitions of Matroids\\[2pt]and Matroid Covering Games}
\author{Pengfei Liu\qquad Han Xiao\qquad Qizhi Fang\\[4pt]
\small School of Mathematical Sciences, Ocean University of China}
\date{}
\makeatletter
\renewcommand{\@maketitle}{%
  \begin{center}
    {\LARGE\@title\par}\vspace{9pt}
    {\large\begin{tabular}{c}\@author\end{tabular}\par}
  \end{center}}
\makeatother
\begin{document}
\maketitle

\begin{abstract}
We give a polynomial time algorithm for computing the prime partition of a matroid and its order at a given rational parameter. The algorithm finds a family of weighted bases and constructs its exchange graph. The strongly connected components give the variable blocks, and reachability gives their order. We apply the algorithm at the fractional covering number to the matroid covering game. The partition characterizes the core when it is nonempty and gives the optimal allocations of the first linear program for the nucleolus. The levels of the order then determine the nucleolus, including when the core is empty.
\end{abstract}

\section{Introduction}
Let $\mathcal M$ be a matroid with ground set $N$ and rank function $r$. Given a rational number $t$, consider the sets $S\subseteq N$ that minimize $t r(S)+|N\setminus S|$. Two elements belong to the same block if they occur in exactly the same minimizing sets. Some elements occur in every minimizer, and some occur in none. For the remaining blocks, we write $Q\preceq Q'$ if every minimizer containing $Q$ also contains $Q'$. We call this partition, together with the order on its variable blocks, the \emph{prime partition at $t$}. Our first problem is to compute it. Our second is to use it to divide the cost of covering a matroid by independent sets.

The minimizing sets are classical in the theory of principal partitions. Narayanan and Vartak~\cite{NV81} studied their construction through matroid union and density. Catlin, Grossman, Hobbs, and Lai~\cite{CGHL92} related principal partitions to strength and fractional arboricity. Fujishige~\cite[Section~3.3]{Fuj09} describes the minimizing lattice at a fixed parameter and the order it induces. These decompositions have several uses. Soto~\cite{Soto13} used the principal sequence to split a matroid secretary problem into problems on uniformly dense minors. Aziz, Lachish, Paterson, and Savani~\cite{ALPS09} used an order on edge blocks to describe optimal strategies in a network wiretap game and compute the nucleolus of a spanning connectivity game.

We compute the partition from exchanges between bases. For the given $t>0$ we find weighted bases of total weight $t$. An arc of the exchange graph records which element can replace which other element in one of these bases. Two graph searches find the elements that belong to every minimizer and those that belong to none. On the remaining elements the strongly connected components give the blocks, and reachability gives their order. We obtain the weighted bases by matroid union. The algorithm runs in polynomial time with an independence oracle, also when $t$ is given in binary.

In the matroid covering game, the players are the elements of a loopless matroid and the cost of a coalition is the minimum number of independent sets needed to cover it. For a graphic matroid, this is the arboricity game. Xiao and Fang~\cite{XF23} used a prime partition to compute its nucleolus when the core is nonempty. Ebert and Ellerbrock~\cite[Theorem~3.4]{EE26} gave a polynomial time algorithm that also covers the empty-core case. Their reduction extends to matroids because the sets coverable in a given number of rounds form a matroid union. Related algorithms use a hierarchy for network strength games~\cite{BB20} and base exchanges for matroid packing games~\cite{LXCF26}.

Computing the nucleolus~\cite{Sch69} usually requires comparing the excesses of exponentially many coalitions. Algorithms for assignment games~\cite{SR94}, balanced simple flow games~\cite{PRB06}, and weighted matching games~\cite{KPT20} exploit the structure of the underlying optimization problem. For the covering game, we use the prime partition at the fractional covering number. We show that the first linear program in the sequence of Maschler, Peleg, and Shapley~\cite{MPS} leaves equal shares within each block and orders those shares between blocks. If several allocations remain, the second program has a unique optimum. The share of an element at this optimum is proportional to the level of its block.

Section~\ref{sec:cover-prelim} defines the prime partition and identifies the parameters for union, packing, and covering. Section~\ref{sec:prime-computation} gives the algorithm. Section~\ref{sec:game} applies it to the matroid covering game.

\section{Prime partitions of a matroid}\label{sec:cover-prelim}
\subsection{Matroid notation}
Let $\mathcal M=(N,\mathcal I)$ be a finite matroid on $n\geq1$ elements. Its rank $r(S)$ is the maximum size of an independent subset of $S$. A \emph{base} is a maximal independent set. Write $\mathcal B$ for the family of bases and $B-u+v$ for $(B\setminus\{u\})\cup\{v\}$. We use the following exchange property. For any two bases $B,B'$, the elements of $B\setminus B'$ can be matched bijectively to those of $B'\setminus B$ so that $B-u+v$ is a base for every matched pair $(u,v)$~\cite[Corollary~39.12a]{Schr03}. For a vector $y$ on $N$, write $y(S)=\sum_{v\in S}y(v)$.

\subsection{The prime partition and its order}
Fix a rational parameter $t$. For $S\subseteq N$, define
\begin{equation}\label{eq:param}
 \phi_t(S)=t r(S)+|N\setminus S|,\qquad
 h(t)=\min_{S\subseteq N}\phi_t(S),\qquad
 \mathcal U_t=\{S\subseteq N:\phi_t(S)=h(t)\}.
\end{equation}
For $t>0$ this is also the minimizing family of $r(S)-\lambda|S|$ at $\lambda=1/t$. Submodularity implies that $\mathcal U_t$ is closed under union and intersection. For $t\leq0$ adding an element strictly decreases $\phi_t$, so $\mathcal U_t=\{N\}$. Set $X_t^-=\bigcap\mathcal U_t$ and $X_t^+=\bigcup\mathcal U_t$.

\begin{definition}\label{def:prime}
On $X_t^+\setminus X_t^-$, put $u\sim_t v$ if $u\in S$ exactly when $v\in S$ for every $S\in\mathcal U_t$. The \emph{prime partition at $t$} consists of these equivalence classes, together with the mandatory part $X_t^-$ and the excluded part $N\setminus X_t^+$ whenever they are nonempty. Its \emph{variable blocks} carry the order
\[
 Q\preceq_t Q'\quad\Longleftrightarrow\quad
 \text{every }S\in\mathcal U_t\text{ containing }Q\text{ also contains }Q'.
\]
\end{definition}

The partition and its order describe every minimizer. A collection of blocks is an \emph{upper set} if it contains every successor of each of its blocks. Each minimizer is $X_t^-$ together with the blocks of an upper set. Conversely, intersect all minimizers containing a given block $Q$. The result is $X_t^-$ together with $Q$ and all its successors. Unions of these sets, together with $X_t^-$ itself, give every upper set and hence every choice allowed by the order.

\subsection{Matroid union}
For a positive integer $j$, the union of $j$ copies of $\mathcal M$, denoted by $\mathcal M^j$, has as independent sets the unions of $j$ independent sets of $\mathcal M$. By the matroid union theorem~\cite[Corollary~42.1a]{Schr03},
\begin{equation}\label{eq:union}
 r_{\mathcal M^j}(S)=\min_{T\subseteq S}\{j r(T)+|S\setminus T|\}.
\end{equation}
Taking $S=N$ gives $r_{\mathcal M^j}(N)=h(j)$. Thus the prime partition at $t=j$ describes all the sets attaining the minimum in the union rank formula.

\subsection{Base packing}
Suppose $r(N)>0$. A base packing is a collection of pairwise disjoint bases. A fractional base packing assigns nonnegative weights to bases so that the total weight on each element is at most one. The maximum total weight is the \emph{strength} $\rho_f$, and the maximum number of disjoint bases is $\lfloor\rho_f\rfloor$~\cite[Chapter~42]{Schr03}. The strength is the largest parameter for which $N$ is a minimizer.
\begin{equation}\label{eq:packing-parameter}
 \rho_f=\min_{T\subseteq N:\,r(T)<r(N)}
          \frac{|N\setminus T|}{r(N)-r(T)},\qquad
 N\in\mathcal U_t\quad\Longleftrightarrow\quad t\leq\rho_f\quad(t>0).
\end{equation}
Indeed, $\phi_t(N)\leq\phi_t(T)$ holds exactly when $t(r(N)-r(T))\leq|N\setminus T|$. At $t=\rho_f$, equality gives the usual packing witnesses
\[
 \{N\setminus T:T\in\mathcal U_{\rho_f}\}
 =\{P\subseteq N:|P|=\rho_f(r(N)-r(N\setminus P))\}.
\]
The nonempty members attain the minimum strength ratio. Taking complements preserves the variable blocks and reverses their order, giving the packing order used in matroid packing games~\cite{LXCF26}.

\subsection{Independent set covering}
Suppose $\mathcal M$ is loopless. A cover of $S$ is a collection of independent sets whose union contains $S$. Write $\gamma(S)$ for its minimum size, with $\gamma(\varnothing)=0$. A fractional cover assigns nonnegative weights to independent sets so that the total weight on each element of $N$ is at least one. Each set can be extended to a base with its weight unchanged. The minimum total weight $a_f$ and the covering number $a^*$ satisfy~\cite{Edm65,Schr03}
\begin{equation}\label{eq:af}
 a_f=\max_{\varnothing\ne T\subseteq N}\frac{|T|}{r(T)},\qquad
 a^*=\gamma(N)=\lceil a_f\rceil .
\end{equation}
Here $\phi_t(\varnothing)\leq\phi_t(T)$ is equivalent to $t r(T)\geq|T|$. Thus $\varnothing\in\mathcal U_t$ precisely when $t\geq a_f$, and
\begin{equation}\label{eq:density-lattice}
 \mathcal F_{\mathrm c}:=\mathcal U_{a_f}=\{T\subseteq N:|T|=a_f r(T)\}.
\end{equation}
The nonempty members of $\mathcal F_{\mathrm c}$ are precisely the sets of maximum density. Their blocks inherit the order $\preceq_{a_f}$. This is the partition used in the covering game.

For example, $U_{2,3}$ has $\rho_f=a_f=3/2$. At this parameter $\varnothing$ and $N$ are the only minimizers, and $N$ is one variable block. At the integer parameters $1$ and $2$, the unique minimizers are $N$ and $\varnothing$, respectively. Rounding gives the correct integral packing and covering numbers, but loses this block.

\section{Computing the prime partition}\label{sec:prime-computation}

For a positive rational $t$, we first compute a family of weighted
bases and then read the prime partition from its exchange graph.
When $t\leq0$, the unique minimizer is $N$ by~\eqref{eq:param}.

\subsection{Computing the weighted bases}

We assign a total weight $t$ to
the bases of $\mathcal M$. The load $c(v)$ of an element is the total
weight of the bases containing it. We choose the weights to maximize
the load covered up to one unit at each element,
\begin{equation}\label{eq:capped-load}
 \max_{\substack{\mu_B\geq0\ (B\in\mathcal B)\\
                  \sum_{B\in\mathcal B}\mu_B=t}}
 L(c),\qquad
 L(c)=\sum_{v\in N}\min\{1,c(v)\},\qquad
 c(v)=\sum_{B\ni v}\mu_B .
\end{equation}

For any $S\subseteq N$, the elements in $S$ contribute at most
$c(S)$ to $L(c)$ and those outside $S$ contribute at most
$|N\setminus S|$. Since each base meets $S$ in at most $r(S)$
elements, $L(c)\leq t r(S)+|N\setminus S|$. Thus $h(t)$ is an upper
bound on $L(c)$.

To find weights attaining this bound, write $t=p/q$ with $p,q$ positive integers. Replace
each element by $q$ parallel copies. Let $N_{[q]}$ be the resulting
ground set and $\pi$ the projection onto $N$. By~\eqref{eq:union},
the maximum number of copies that can be partitioned into $p$
independent sets is
\[
 \min_{Y\subseteq N_{[q]}}\{p\,r(\pi Y)+q|N|-|Y|\}
 =q\min_{S\subseteq N}\{(p/q)r(S)+|N\setminus S|\}
 =q h(t).
\]
The first minimum is attained on whole classes of copies, since adding
a missing copy from a represented class preserves rank. Project the
$p$ independent sets of a maximum union to $N$, and extend each to a
base, using an arbitrary base for an empty class. Give each base
weight $1/q$. Before extension every load is at most one, and the
sum of the loads is $h(t)$. Extending the sets can only increase $L(c)$,
so the resulting weighting attains $h(t)$.

For a parameter given in binary, we can obtain these weights using
at most two matroid unions of polynomial size. Put
\[
 \mathcal T=\{a/b:a,b\in\mathbb Z,\ 0\leq a\leq n,\ 1\leq b\leq n\}.
\]
For $0<t\leq n$, each point where two distinct objective lines
$\phi_t(S)$ and $\phi_t(T)$ cross belongs to $\mathcal T$. Indeed,
their slopes are ranks and their intercepts are cardinalities, all
between zero and $n$. Hence the lower envelope $h$ is affine on the
closed interval between any two consecutive entries of $\mathcal T$.

At a positive entry $a/b$ in lowest terms, obtain an optimal weighting
by the union construction with $a$ independent sets on $bn$ copies.
At zero use the empty weighting. If $t$ is an entry, this gives the
required weights directly. Otherwise, for $0<t<n$, take the
consecutive entries $\alpha<t<\beta$ and set
$\lambda=(t-\alpha)/(\beta-\alpha)$. Multiply the optimal weights at
$\alpha$ by $1-\lambda$, those at $\beta$ by $\lambda$, and add the
weights on each base. Their total is $t$. Writing $c_\alpha,c_\beta$
for the endpoint loads, the new load is
$c=(1-\lambda)c_\alpha+\lambda c_\beta$. Concavity of $L$ gives
\[
 L(c)\geq(1-\lambda)L(c_\alpha)+\lambda L(c_\beta)
       =(1-\lambda)h(\alpha)+\lambda h(\beta)=h(t).
\]
The upper bound $L(c)\leq h(t)$ now gives equality, so the
weights solve~\eqref{eq:capped-load} at the prescribed parameter.

If $t>n$, let $P$ be the set of nonloops. For each $v\in P$, give
unit weight to a base containing $v$. Put the remaining weight
$t-|P|$ on any base, using the empty base in rank zero. This gives
$L(c)=|P|$, and taking $S=N\setminus P$ gives $\phi_t(S)=|P|$.
Thus the weighting is again optimal.

The list $\mathcal T$ contains $O(n^2)$ fractions. At each endpoint
the union algorithm uses at most $n$ colors on at most $n^2$
copies~\cite[Chapter~42]{Schr03}. An independence test in the parallel
extension checks that the elements have distinct projections and
queries independence of those projections in $\mathcal M$. Projection
and greedy extension give at most $2n$ bases. The number of oracle
calls is polynomial in $n$, and each weight has encoding length
polynomial in $n$ and in the encoding length of $t$.

\subsection{From base exchanges to the prime partition}

Let $B_1,\ldots,B_s$ be the bases of positive weight in an optimal
solution of~\eqref{eq:capped-load}. The following steps return the
prime partition at $t$ and its order.
\begin{enumerate}
\item Construct a directed graph $G$ on $N$. For each $B_i$, insert
an arc $v\to u$ if $v\notin B_i$, $u\in B_i$, and $B_i-u+v$ is a base.
The arc points from the entering element to the leaving element.
\item Compute the strongly connected components of $G$ and contract
each component to a vertex. In this acyclic graph, mark every component
reachable from a component containing an element of load less than one. Their union $A$ is
the mandatory part. Mark every component that can reach one
containing an element of load greater than one. Their union $D$ is the excluded part.
Both searches include their starting components.
\item Keep each remaining component as a separate block. A block $Q$
precedes $Q'$ if there is a directed path from $Q$ to $Q'$.
Return these ordered blocks together with $A$ and $D$, omitting
either of the last two parts when it is empty.
\end{enumerate}
Write $V=N\setminus(A\cup D)$ for the union of the variable blocks.

\begin{theorem}\label{thm:graph}
For any optimal weighting in~\eqref{eq:capped-load}, the exchange
graph gives
$X_t^-=A$ and $X_t^+=N\setminus D$. For two variable blocks $Q,Q'$,
we have $Q\preceq_t Q'$ exactly when $Q'$ is reachable from $Q$.
The minimizers are precisely the unions of $A$ with upper sets of
these blocks.
\end{theorem}
\begin{proof}
For a weighting with positive weights $\mu_i$ on $B_i$ and any
$S\subseteq N$, expand $\phi_t(S)-L(c)$ to obtain
\begin{equation}\label{eq:load-gap}
\begin{aligned}
 \phi_t(S)-L(c)
 ={}&\sum_{i=1}^s\mu_i\bigl(r(S)-|B_i\cap S|\bigr)\\
 &+\sum_{v\in S}(c(v)-1)^+
  +\sum_{v\notin S}(1-c(v))^+ ,
\end{aligned}
\end{equation}
where $z^+=\max\{z,0\}$. Since $L(c)=h(t)$, a set $S$ is a minimizer precisely when all three
terms vanish.

Call $S$ closed if no arc leaves $S$.
The exchange graph gives the equivalence
\begin{equation}\label{closed-rank}
 S\text{ is closed in }G
 \quad\Longleftrightarrow\quad
 |B_i\cap S|=r(S)\quad(1\leq i\leq s).
\end{equation}
An arc leaving $S$ exchanges an element outside $S$ for one in $S$,
increasing the intersection of its base with $S$. Conversely, if
$|B_i\cap S|<r(S)$, take $v\in S\setminus B_i$ for which
$(B_i\cap S)\cup\{v\}$ is independent. The fundamental circuit of $v$
in $B_i\cup\{v\}$ contains some $u\in B_i\setminus S$. The exchange
$B_i-u+v$ gives an arc leaving $S$.

By~\eqref{eq:load-gap} and~\eqref{closed-rank}, a set minimizes
$\phi_t$ precisely when it is closed, contains every element of load
less than one, and avoids every element of load greater than one.
Consequently every minimizer contains $A$ and avoids $D$, so
$A\cap D=\varnothing$. The sets $A$ and $N\setminus D$ are themselves
closed and satisfy the two load conditions. They are the smallest
and largest minimizers.

For $v\in V$, let $R(v)$ be the set of vertices reachable from $v$,
including $v$. This set avoids $D$, so $A\cup R(v)$ is the smallest
minimizer containing $v$. For $v,w\in V$, membership of $v$ therefore
forces membership of $w$ exactly when $v$ reaches $w$. A path between variable vertices stays in $V$, since entering $D$
would put $v$ in $D$ and entering the closed set $A$ would put $w$ in
$A$. Mutual reachability now gives the variable blocks. The upper sets
of these components give all minimizers when added to $A$.
\end{proof}

For $s$ bases, constructing the graph takes at most
$s\,r(N)(n-r(N))$ independence queries. Computing the components,
contracting them and running the two searches take $O(n+|E(G)|)$ time.
Together with the computation of the weights, this gives a polynomial
time algorithm for the ordered prime partition at the given $t$.

\begin{example}\label{ex:interior}
Let $C=\{c_1,c_2,c_3\}$, $H=\{h_1,h_2\}$ and $L=\{\ell_1,\ell_2\}$.
Take the direct sum of $U_{1,3}$ on $C$, a loopless rank-two matroid
on $H\cup L$ whose only parallel pair is $H$, and a coloop $g$.
Consider the partition at $t=2$. Give unit weight to each of the bases
\[
 \{c_1,h_1,\ell_1,g\},\qquad \{c_2,h_2,\ell_2,g\}.
\]
These weights are optimal since $L(c)=7=\phi_2(C)$.
The forward search from $c_3$, the only element of load zero, reaches
all of $C$. The backward search from $g$, the only element of load two,
reaches only $g$. On $H\cup L$ the exchanges in both directions within
each pair give two strongly connected components, $H$ and $L$.
An unused element of $L$ can replace the element of $H$ in either base,
giving $L\prec_2 H$. The reverse exchange would put both parallel
elements of $H$ in a base. Thus the minimizing sets are
$C$, $C\cup H$ and $C\cup H\cup L$. At the covering parameter
$a_f=3$, the only minimizing sets are $\varnothing$ and $C$.
\end{example}

\paragraph{Using one graph for several parameters.}
The same exchange graph can serve all positive parameters if the bases
are chosen from a minimum norm base. Let $b$ minimize the Euclidean norm
over the convex hull of the base incidence vectors. Write
$b=\sum_i\theta_i\mathbf 1_{B_i}$, with $\theta_i>0$ and $\sum_i\theta_i=1$,
and form the exchange graph of the $B_i$.

\begin{corollary}\label{thm:universal-graph}
At each positive rational $t$ the mandatory elements are those with $t b(v)<1$, and the
excluded elements are those with $t b(v)>1$. On the level $t b(v)=1$,
the strongly connected components are the variable blocks, ordered by
reachability.
\end{corollary}
\begin{proof}
An arc $v\to u$ allows weight to move from $B_i$ to $B_i-u+v$.
The squared norm is minimized at $b$, so its derivative in the
direction $\mathbf 1_{\{v\}}-\mathbf 1_{\{u\}}$ gives $b(v)\geq b(u)$.
Hence $L=\{v:t b(v)<1\}$ and $U=\{v:t b(v)\leq1\}$ are both closed
under outgoing arcs. By~\eqref{closed-rank}, $b(L)=r(L)$ and
$b(U)=r(U)$. Give $B_i$ weight $t\theta_i$. Then $c(v)=t b(v)$ and
\[
 \sum_{v\in N}\min\{1,c(v)\}
     =t b(L)+|N\setminus L|
     =t r(L)+|N\setminus L|.
\]
The left side is a lower bound on every $\phi_t(S)$, so $L$ is a
minimizer and these weights are optimal. In Theorem~\ref{thm:graph}
the forward search stays in $L$
and the backward search stays outside $U$. The remaining graph is exactly
the one induced by the level $t b(v)=1$.
\end{proof}

Fujishige's decomposition algorithm computes $b$ in polynomial
time~\cite{Fuj80,Fuj09}. A base decomposition can be obtained by a
minimum fractional cover with demand $b$~\cite[Corollary~42.6a]{Schr03}.
For $r(N)>0$ the optimum is one since $b$ belongs to the base polytope
and $b(N)=r(N)$. Equality forces all loads to equal $b$ and every
set of positive weight to be a base. In rank zero the only base is empty.
For a loopless matroid, comparing the two threshold tests with
\eqref{eq:packing-parameter} and~\eqref{eq:af} gives
$\rho_f=1/\max_v b(v)$ and $a_f=1/\min_v b(v)$~\cite{TP25}.

\section{Matroid covering games}\label{sec:game}

\subsection{Cooperative games}

A cooperative game with transferable utility assigns a value
$\gamma(S)$ to each coalition $S$ of a finite player set $N$, with
$\gamma(\varnothing)=0$. In a cost game, $\gamma(S)$ is the cost
that $S$ would incur on its own. An allocation $x$ divides the total
cost $\gamma(N)$ among the players. In this paper the shares are
nonnegative, so the set of allocations is
$X_0=\{x\in\mathbb R_{\geq0}^N:x(N)=\gamma(N)\}$.
The \emph{excess} of $S$ at $x$ is $e(S,x)=\gamma(S)-x(S)$.
An allocation belongs to the \emph{core} $\mathcal C$ if every
coalition pays at most its own cost, that is, if every excess is nonnegative.

Let $\theta(x)$ list the coalition excesses in nondecreasing order.
The \emph{nucleolus} $\eta$ is the unique allocation in $X_0$ that
lexicographically maximizes $\theta(x)$~\cite{Sch69}. We include
the empty and grand coalitions, whose excess is zero at every allocation.

\subsection{The matroid covering game and the linear programming sequence}
\label{sec:game-sequence}

Let $\mathcal M=(N,\mathcal I)$ be a loopless matroid. In its
\emph{covering game}, each element is a player and the cost $\gamma(S)$
is the minimum number of independent sets needed to cover $S$.
By~\eqref{eq:af}, the total cost to allocate is
$a^*=\gamma(N)=\lceil a_f\rceil$.

We use the sequence of linear programs of Maschler, Peleg, and
Shapley~\cite{MPS}, also used for matroid packing games~\cite{LXCF26}.
For a nonempty set $X$ of allocations, write
\[
 \operatorname{fix}(X)=
 \{S\subseteq N:x(S)=y(S)\text{ for every }x,y\in X\}.
\]
Start with $\mathcal V_0=2^N$. At stage $k$ solve
\begin{equation}\label{eq:game-lp}
 \begin{aligned}
  (\mathrm{LP}_k)\qquad \max_{x,\varepsilon}\quad &\varepsilon\\
  \text{subject to}\quad &x\in X_{k-1},\\
                         &\gamma(S)-x(S)\geq\varepsilon
                               \quad(S\in\mathcal V_{k-1}).
 \end{aligned}
\end{equation}
Let $\varepsilon_k$ be the optimal value and $X_k$ the set of optimal
allocations. The first optimal set $X_1$ is the \emph{least core}.
If $X_k$ is a singleton, its member is the nucleolus and the sequence
stops. Otherwise set $\mathcal V_k=2^N\setminus\operatorname{fix}(X_k)$
and continue. The next program maximizes the smallest excess among
the coalitions whose excess can still vary on $X_k$.

\subsection{The first linear program and the core}\label{sec:game-first}

For the covering game the first program is

\begin{equation}\label{eq:lp1}
 \begin{aligned}
  (\mathrm{LP}_1)\qquad \max_{x,\varepsilon}\quad &\varepsilon\\
  \text{subject to}\quad &x\geq 0,\quad x(N)=a^*,\\
                         &x(S)+\varepsilon\leq\gamma(S)\quad(S\subseteq N).
 \end{aligned}
\end{equation}
The constraint for $N$ gives $\varepsilon_1\leq0$. Hence the core is
nonempty precisely when $\varepsilon_1=0$, in which case
$X_1=\mathcal C$.

Run the algorithm of Section~\ref{sec:prime-computation} at $t=a_f$.
To find $a_f$, search the positive fractions in $\mathcal T$ for the
smallest $t$ with $\varnothing\in\mathcal U_t$. At this parameter
$L(c)=h(a_f)=\phi_{a_f}(\varnothing)=|N|$, so every element has load
at least one. The weighted bases computed in Section~\ref{sec:prime-computation}
therefore form a minimum fractional cover, and the mandatory part is empty.
Write $W=X^+_{a_f}$ for the union of the variable
blocks, $\mathcal Q^c$ for these blocks and $\preceq$ for their order.
All excluded elements form a single block $N\setminus W$ when this set is nonempty.

The dual of the fractional covering problem maximizes $y(N)$ subject to
$y\geq0$ and $y(B)\leq1$ for every base $B$. Its set of optima is
\begin{equation}\label{eq:programs}
 \mathcal D(\mathcal M)=\{y\geq0:y(N)=a_f,\ y(B)\leq1\ (B\in\mathcal B)\}.
\end{equation}
For a prescribed total $b>0$, let $F(b)$ consist of the vectors satisfying

\begin{equation}\label{cover-order-face}
\begin{aligned}
 &x\geq 0,\qquad x(N)=b,\qquad x(v)=0 &&(v\in N\setminus W),\\
 &x(v)=x_Q &&(v\in Q,\ Q\in\mathcal Q^c),\\
 &x_Q\leq x_{Q'} &&(Q\preceq Q').
\end{aligned}
\end{equation}

Here $x_Q$ denotes the common share of the elements in $Q$.

\begin{theorem}\label{cover-core}
The core is nonempty if and only if $a_f$ is an integer.
In this case $\mathcal C=F(a^*)$.
\end{theorem}
\begin{proof}
Every base has cost one. Conversely, if $x\geq0$ satisfies
$x(B)\leq1$ for every base, then $x(I)\leq1$ for every independent
set $I$, since $I$ extends to a base. Partitioning $S$ into
$\gamma(S)$ independent sets then gives $x(S)\leq\gamma(S)$.
The largest total allowed by the base inequalities is $a_f$.
Since $a^*=\lceil a_f\rceil$, the required total is attainable
precisely when $a^*=a_f$, and then $\mathcal C=\mathcal D(\mathcal M)$.
It remains to show that $\mathcal D(\mathcal M)=F(a_f)$.

Fix any minimum fractional base cover with positive weights $\mu_i$ on
the bases $B_i$ and loads $c(v)=\sum_{i:v\in B_i}\mu_i$. Every load is
at least one. Write $\widetilde N_0=\{v:c(v)>1\}$. By
Theorem~\ref{thm:graph}, the vertices that can reach $\widetilde N_0$
are exactly $N\setminus W$. For any dual optimum $y$, the base
inequalities imply
\[
 a_f=y(N)\leq\sum_{v\in N}c(v)y(v)
   =\sum_i\mu_i y(B_i)\leq\sum_i\mu_i=a_f .
\]
Thus $y(v)=0$ for $v\in\widetilde N_0$ and $y(B_i)=1$ for each $i$.
An arc $v\to u$ means that $B_i-u+v$ is a base for some $B_i$.
Comparing the two base weights gives $y(v)\leq y(u)$.
It follows that $y$ vanishes on $N\setminus W$, is constant on each
variable block, and is nondecreasing along the order.

Conversely, let $y\in F(a_f)$. Write its positive coordinate values
as $0<t_1<\cdots<t_m$ and put $t_0=0$. Every nonempty level set
$S_i=\{v:y(v)\geq t_i\}$ is an upper set of blocks, so
$|S_i|=a_f r(S_i)$. For any base $B$ we have
\[
 y(B)=\sum_{i=1}^{m}(t_i-t_{i-1})|B\cap S_i|
 \leq \sum_{i=1}^{m}(t_i-t_{i-1})r(S_i)
 =\frac{y(N)}{a_f}=1 .
\]
This proves $\mathcal D(\mathcal M)=F(a_f)$.
\end{proof}

Two consequences of the proof will be used below. Rescaling the
dual optimal face gives $F(b)=(b/a_f)\mathcal D(\mathcal M)$.
Every base of positive weight in a minimum fractional cover has
the same $x$-weight $b/a_f$ throughout $F(b)$.

If $a^*=1$, then $\mathcal M$ is free and every nonempty coalition
costs one. In this case $X_1=\mathcal C=X_0$ and $\varepsilon_1=0$.
To determine the first optimal face when $a^*\geq2$, we consider the
coalitions that can be covered in $a^*-1$ rounds. They are the
independent sets of the corresponding union matroid.

\begin{lemma}\label{cover-union-density}
For every integer $1\leq j<a_f$, the fractional covering value of
$\mathcal M^j$ is $a_f/j$. The two matroids have the same sets of
maximum density, the same covering prime blocks and the same order.
Moreover, $\mathcal D(\mathcal M^j)=j^{-1}\mathcal D(\mathcal M)$.
\end{lemma}
\begin{proof}
The matroid union rank formula gives
\[
 r_{\mathcal M^j}(S)=\min_{T\subseteq S}
 \bigl\{|S\setminus T|+j r(T)\bigr\}.
\]
Since $a_f r(T)\geq |T|$, every expression in braces is at least
\[
 (j/a_f)|S|+(1-j/a_f)|S\setminus T| .
\]
Since $j<a_f$, the lower bound $(j/a_f)|S|$ can be attained only
at $T=S$. It is attained there exactly when $|S|=a_f r(S)$.
At least one nonempty set has this property. Thus the largest density
in $\mathcal M^j$ is $a_f/j$, attained at exactly the same sets as in
$\mathcal M$. These sets define the prime blocks and their order, so
the two dual optimal faces in~\eqref{cover-order-face} differ only
in their totals.
\end{proof}

\begin{theorem}\label{cover-least-core}
The optimal value and the set of optimal allocations of $(\mathrm{LP}_1)$ are
\begin{equation}\label{cover-least-core-value}
 \varepsilon_1=(a^*-1)\left(1-\frac{a^*}{a_f}\right),
 \qquad X_1=F(a^*).
\end{equation}
\end{theorem}
\begin{proof}
For $a^*=1$ this follows from the preceding description of the core.
Suppose $a^*\geq2$.
Put $k=a^*-1$ and $d=a^*/a_f$. If $x\in F(a^*)$, then every base
has weight at most $d$. A coalition of cost $j\leq k$ is a disjoint
union of $j$ independent sets, and its excess is at least
$j(1-d)\geq k(1-d)$. A coalition of cost $a^*$ has excess
$x(N\setminus S)\geq0$. Thus $F(a^*)$ attains a minimum excess at
least $k(1-d)$.

For the upper bound, put $\mathcal H=\mathcal M^k$. Every base $K$
of $\mathcal H$ is nonempty and proper, since $N$ requires $k+1$
independent sets. Suppose that $x$ has minimum excess at least
$\varepsilon$. Then $x(K)\leq\gamma(K)-\varepsilon\leq k-\varepsilon$.
Average these inequalities with the weights of an optimal fractional
base cover of $\mathcal H$. Every element is covered to weight at
least one. By nonnegativity of $x$ and Lemma~\ref{cover-union-density},
\[
 a^*=x(N)\leq (a_f/k)(k-\varepsilon).
\]
It follows that $\varepsilon\leq k(1-d)$. At equality, the base
inequalities $x(K)\leq kd$ and the total $x(N)=a^*$ place
$x/(kd)$ in $\mathcal D(\mathcal H)$.
Lemma~\ref{cover-union-density} gives $x/d\in\mathcal D(\mathcal M)$.
This proves both claims.
\end{proof}

Every allocation in $X_1$ is individually rational. Indeed, for a
block $Q$, let $U$ be the union of $Q$ and its strict successors.
Both $U$ and $U\setminus Q$ are tight in~\eqref{eq:density-lattice},
so $|Q|/a_f=r(U)-r(U\setminus Q)$ is a positive integer. Hence
$|Q|\geq\lceil a_f\rceil=a^*$ and $x_Q\leq a^*/|Q|\leq1=\gamma(\{v\})$
for $v\in Q$. The remaining players pay zero.

\subsection{The nucleolus}\label{sec:game-nucleolus}

By Theorem~\ref{cover-least-core}, players in the same block pay the
same share throughout $X_1$, and the shares are nondecreasing along
the block order. If there is only one block in $W$, its share is fixed
by the total cost. Otherwise let
$\mathcal V=2^N\setminus\operatorname{fix}(X_1)$.
The second program in~\eqref{eq:game-lp} is

\begin{equation}\label{eq:lp2}
 \begin{aligned}
   (\mathrm{LP}_2)\qquad \max_{x,\varepsilon}\quad &\varepsilon\\
   \text{subject to}\quad &x\in X_1,\\
                          &x(S)+\varepsilon\leq\gamma(S)\quad(S\in\mathcal V).
 \end{aligned}
\end{equation}
Write $Q\lessdot Q'$ if $Q\prec Q'$ and no block lies strictly
between them. The next lemma reduces the coalition constraints to a
lower bound on each block share and on the difference between
consecutive comparable shares. Write $z=\varepsilon-\varepsilon_1$.

\begin{lemma}\label{cover-lp-equivalence}
Suppose $a^*\geq2$ and there are at least two variable blocks.
For $x\in X_1$ and $z\geq0$, the coalition constraints of
$(\mathrm{LP}_2)$ at $\varepsilon=\varepsilon_1+z$ are equivalent to
\begin{equation}\label{cover-local-lp}
 x_Q\geq z\quad(Q\in\mathcal Q^c),\qquad
 x_{Q'}-x_Q\geq z\quad(Q\lessdot Q').
\end{equation}
\end{lemma}
\begin{proof}
First note that for any two distinct blocks the shares $x_Q$ and
the differences $x_{Q'}-x_Q$ vary on $X_1$. To see this, choose
positive shares strictly increasing along a linear extension and
normalize their total to $a^*$. Any sufficiently small change of
these shares that preserves the total is feasible in~\eqref{cover-order-face}.

Suppose every coalition in $\mathcal V$ has excess at least
$\varepsilon_1+z$. Let
$k=a^*-1$, $d=a^*/a_f$ and $\mathcal H=\mathcal M^k$. Fix a
minimum fractional base cover of $\mathcal H$. For every
positive-weight base $K$ in this cover, complementary slackness
applied to $x/(kd)\in\mathcal D(\mathcal H)$ gives
\[
 x(K)=kd=k-\varepsilon_1\qquad(x\in X_1).
\]
For $v\in Q$, choose such a $K$ containing $v$. The weight of $K-v$
is $kd-x_Q$ and varies on $X_1$. Since $\gamma(K-v)\leq k$,
$\varepsilon_1+z\leq e(K-v,x)\leq k-kd+x_Q=\varepsilon_1+x_Q$.

For $Q\lessdot Q'$, Lemma~\ref{cover-union-density} gives the same
two consecutive blocks in $\mathcal H$. In its exchange graph a
path from $Q$ to $Q'$ stays in $W$ and cannot meet an intermediate
component. Hence there is an arc $u\to v$ with $u\in Q$ and $v\in Q'$.
There is consequently a positive-weight base $K$ of $\mathcal H$
with $v\in K$, $u\notin K$, and $K-v+u$ a base of $\mathcal H$.
This last coalition has nonconstant weight
$kd-x_{Q'}+x_Q$ and cost at most $k$. Its inequality gives
$x_{Q'}-x_Q\geq z$.

Conversely, suppose~\eqref{cover-local-lp} holds and let
$S\in\mathcal V$. If $\gamma(S)=a^*$, then $e(S,x)=x(N\setminus S)$.
Since this quantity varies on $X_1$, the complement contains an
element of $W$. Its coordinate is at least $z$, and hence
$e(S,x)\geq z\geq\varepsilon_1+z$.

If $j=\gamma(S)\leq k$, partition $S$ into independent sets
$I_1,\ldots,I_j$, and extend $I_h$ to a base $B_h'$ of $\mathcal M$.
Fix one positive-weight base $B$ of an optimal fractional base
cover of $\mathcal M$, so $x(B)=d$ on $X_1$. For each $h$, the
base-exchange matching pairs $u\in B\setminus B'_h$ with
$v\in B'_h\setminus B$ such that $B-u+v$ is a base. Denote the
collection of pairs by $P_h$. Subtracting $x(S)$ from $j$ gives
\begin{equation}\label{cover-coalition-decomposition}
 e(S,x)=j(1-d)+
 \sum_{h=1}^j\sum_{(u,v)\in P_h}\bigl(x(u)-x(v)\bigr)
       +\sum_{h=1}^j x(B'_h\setminus I_h).
\end{equation}
Each matched pair supplies an arc $v\to u$ in the covering graph,
so $x(v)\leq x(u)$ throughout $X_1$. A nonconstant difference
therefore has either $v\notin W$, $u\in W$, or two distinct,
comparable blocks as endpoints. In the first case its value is at
least $z$ by the coordinate bound. In the second case, summing
along a chain of cover relations gives the same bound.
All coordinates in the last sum are nonnegative, and each
nonconstant coordinate is again at least $z$.
Since $S\in\mathcal V$, some term after the constant in
\eqref{cover-coalition-decomposition} must vary on $X_1$. Thus
$e(S,x)\geq j(1-d)+z\geq\varepsilon_1+z$.
\end{proof}

If a block share must exceed each predecessor's share by at least
$z$, then a chain of $i$ blocks gives a lower bound of $iz$ at its
last block. This leads to the following levels of the block order.

\begin{definition}\label{def:game-hierarchy}
Let $\mathcal Q_1^c$ consist of the minimal blocks of $(\mathcal Q^c,\preceq)$.
For each $i>1$, let $\mathcal Q_i^c$ consist of the minimal blocks after
the first $i-1$ levels have been removed. Denote the number of levels by
$\ell$ and set $n_i=\sum_{Q\in\mathcal Q_i^c}|Q|$. A block belongs to
level $i$ exactly when a longest chain ending at that block has $i$ members.
\end{definition}

\begin{theorem}\label{cover-nucleolus}
If $X_1$ is a singleton, its member is the nucleolus.
Otherwise $(\mathrm{LP}_2)$ has a unique optimal allocation, which is
the nucleolus. In both cases its coordinates are
\begin{equation}\label{cover-nucleolus-formula}
 \eta(v)=
 \begin{cases}
 \displaystyle\frac{i\,a^*}{\sum_{h=1}^{\ell} h n_h},
      &v\in Q,\ Q\in\mathcal Q^c_i,\\[6pt]
 0,   &v\in N\setminus W .
 \end{cases}
\end{equation}
\end{theorem}

\begin{proof}
If there is only one variable block, its share is $a^*/|W|$ and all
other elements receive zero, as claimed. If $\mathcal M$ is free and
$|N|\geq2$, its blocks are singletons in one level and $X_1=X_0$.
For every nonempty proper $S$ we have $e(S,x)=x(N\setminus S)$.
The minimum over these coalitions is $\min_{v\in N}x(v)$, attained
by the complement of a singleton. Its unique maximizer has
$x(v)=1/|N|$ for all $v$, as in~\eqref{cover-nucleolus-formula}.
Now suppose $a^*\geq2$ and there are at least two blocks.
By Lemma~\ref{cover-lp-equivalence}, the second program maximizes
$z$ subject to $x\in X_1$ and~\eqref{cover-local-lp}. The value
$z=0$ is feasible.
Induction along a longest chain gives $x_Q\geq iz$ for every block
at level $i$. Summing over active elements yields
\[
 a^*=x(N)\geq z\sum_{i=1}^{\ell} i n_i .
\]
Set $z=a^*/\sum_i i n_i$ and assign $iz$ to each player on level $i$.
This allocation has the required total and satisfies~\eqref{cover-local-lp},
since each cover relation increases the level by at least one.
Every optimal allocation must attain $x_Q=iz$ on every block of level
$i$ to attain equality in the sum. Thus the displayed allocation is the
unique optimum of $(\mathrm{LP}_2)$ and is the nucleolus.
\end{proof}

To compute $\eta$, take the exchange graph at $t=a_f$ computed in
Section~\ref{sec:prime-computation}. Assign zero to $N\setminus W$.
Process the components of the graph on $W$ in topological order,
giving a source level one and each other component one more than
the largest level of its immediate predecessors. Counting the players
at each level then gives~\eqref{cover-nucleolus-formula}. Once the
condensation graph is available, these steps take time linear in its
size plus $|N|$.

\paragraph{A two-level example.}
Let $H$ and $L$ each have $m$ elements, where $m=4$ or $5$.
In one round we can cover at most four elements of $H\cup L$, at most
two of them from $H$. These are the independent sets of the rank-four
truncation of $U_{2,m}$ on $H$ in direct sum with the free matroid on
$L$. Add a coloop $g$, which can join any round.

We have $a_f=m/2$. Indeed, $H$ has rank two and $H\cup L$ has rank
four, so both have density $m/2$. For $S\subseteq H\cup L$ put
$h=|S\cap H|$ and $l=|S\cap L|$. If $r(S)=4$, equality in
$h+l\leq (m/2)r(S)$ requires $S=H\cup L$. If $r(S)<4$,
then $r(S)=\min\{2,h\}+l$, and equality for a nonempty $S$ requires
$h=m$ and $l=0$. Adding $g$ increases size and rank by one each,
so every set containing $g$ has density strictly below $m/2$.
The only nonempty maximizers are therefore
$H$ and $H\cup L$.

The exchange graph gives the same order. Label the elements of $H$
and $L$ cyclically and give each base
$\{h_i,h_{i+1},l_i,l_{i+1},g\}$ weight $1/2$. These bases have
total weight $m/2$. Each element in the two classes has load one,
and $g$ has load $m/2$. Replacing an $H$ element by an unused $L$
element gives an arc from $L$ to $H$. The reverse exchange would use
three elements of $H$. Thus $L$ is at level one, $H$ is at level two,
and $g$ receives zero.

Write $p$ and $q$ for the shares on $L$ and $H$. The first program
leaves the segment $m(p+q)=a^*$, $0\leq p\leq q$. The second sets
$q=2p$. Table~\ref{tab:two-level} gives the results for both values of $m$.
\begin{table}[htbp]
\centering
\caption{Block shares at the nucleolus for the two-level example.}
\label{tab:two-level}
\smallskip
\begin{tabular}{c c c c c c c}
\hline
$m$ & $a_f$ & $a^*$ & Core & $\varepsilon_1$ & $p$ on $L$ & $q$ on $H$\\
\hline
$4$ & $2$   & $2$ & nonempty & $0$    & $1/6$ & $1/3$\\
$5$ & $5/2$ & $3$ & empty    & $-2/5$ & $1/5$ & $2/5$\\
\hline
\end{tabular}
\end{table}

\end{document}